\documentstyle[preprint,aps,prl,floats,epsf]{revtex}
\newcommand{\be}{\begin{equation}}
\newcommand{\ee}{\end{equation}}
\newcommand{\bea}{\begin{eqnarray}}
\newcommand{\eea}{\end{eqnarray}}
\newcounter{dafigcounter}
\newcommand{\pfig}[3]{
 \refstepcounter{dafigcounter}
 \begin{minipage}[t]{#2}
  \begin{center}
   {\epsfxsize=#2 \mbox{\epsffile{#1.eps}}}
  \end{center}
  \label{#1}
  \small \bf Fig.~\thedafigcounter\rm\ #3
 \end{minipage}
}
\tightenlines
\begin{document}
\draft

%
\title{Gluon Pair Production From Space-Time Dependent 
Chromofield}
\author{Gouranga C. Nayak and Walter Greiner}  
\address
{\small\it{Institut f\"ur Theoretische Physik,
Robert-Mayer Str. 8-10,
Johann Wolfgang Goethe-Universit\"at,
60054 Frankfurt am Main, Germany}}
\maketitle

\begin{abstract} 

We compute the probabilty for the processes $A \rightarrow q\bar q, gg$ 
via vacuum polarization in the presence of a classical space-time 
dependent non-abelian field $A$ by applying the 
background field method of QCD. 
The processes we consider are leading order in $gA$
and are simillar to $A \rightarrow e^+e^-$ in QED. 
Gluons are treated like a matter field and gauge transformations
of the quantum gluonic field are different from those of the classical
chromofield. To obtain the correct physical polarization of gluons
we deduct the corresponding ghost contributions. 
We find that the expression for the probability of the leading
order process $A \rightarrow gg$ is transverse with respect to 
the momentum of the field. We observe that the contributions 
from higher order processes to gluon pair production 
need to be added to the this leading order process. 
Our result presented here is a part of the expression for the
total probability for gluon pair production from a space-time dependent
chromfield.
The result for $q\bar q$ pair production is similar to that of the
$e^+e^-$ pair production in QED. Quark and gluon production 
from a space-time dependent chromofield will
play an important role in the study of the production and equilibration
of the quark-gluon plasma in ultra relativistic heavy-ion collisions.

\end{abstract}  

\vspace{0.5cm}


\section{Introduction}

Over the years there have been several investigations 
on the production of charged
particles from a classical electro-magnetic field, a phenomenon which was 
discovered nearly fifty years ago by Schwinger \cite{schwinger}. By now, the 
production of electron-positron pairs from the abelian field is extensively 
studied both theoretically and experimentally \cite{all}. The subject 
of quark/anti-quark and gluon pair production from the non-abelian 
field is relatively new and is not fully solved. It might be important 
for the production of the 
quark-gluon plasma (QGP) in the laboratory by high energy 
heavy-ion collisions. Lattice QCD predicts the existence of such a state of
matter at high temperatures ($\sim$ 200 MeV) and densities \cite{lattice}. 
In high energy heavy-ion collisions at RHIC and LHC \cite{rhic} the 
receding nuclei might produce a strong chromofield \cite{nayak,mclerran} 
which would then 
polarize the QCD vacuum and produce quark/anti-quark pairs and gluons. 
These produced quarks and gluons collide with each other to form a thermalized 
quark-gluon plasma. The space-time evolution of the quark-gluon plasma in the
presence of a background chromofield is studied by solving relativistic 
non-abelian transport equation of quarks and gluons with all the dynamical
effects taken into account. As color is a dynamical variable in the non-abelian
theory, the relativistic non-abelian transport equation
for quark and gluon is \cite{nayak,heinz} 
\begin{equation}
\left[ p_{\mu} \partial^\mu + g Q^a F_{\mu\nu}^a p^\nu 
\partial^\mu_p
+ g f^{abc} Q^a A^b_\mu
p^{\mu} \partial_Q^c \right]  f(x,p,Q)=C(x,p,Q)+S(x,p,Q).
\label{trans1}
\end{equation}
Here $f(x,p,Q)$ is the single-particle distribution function of the parton
in the 14 dimensional extended phase space which includes coordinate, momentum
and color in SU(3). The first term on the LHS of Eq.\ (\ref{trans1})
corresponds to the usual convective flow, the second term is the
non-Abelian version of the Lorentz-force term and the third term
corresponds to the precession of the color charge, as described by
Wong's equation \cite{wong}. $C$ and $S$ on the RHS of Eq.\
(\ref{trans1}) are the collision and the source terms, respectively.
Note that there are separate transport equations for quarks,
antiquarks and gluons \cite{nayak,heinz}. The source term $S$ contains
the detailed information about the production of quarks and 
gluons from the chromofield. It is defined as the probability W for the
parton production per unit time per unit volume of the phase space.
Hence the space-time evolution of the quark-gluon plasma in a high
energy heavy-ion collision crucially depends on how quarks and gluons 
are produced from the chromofield \cite{nayak}. 

The production of fermion pairs from the field is studied in two different 
cases: 1) from a constant, uniform electric field and 2) 
from a space-time dependent field. The fermion pair production from a
constant, uniform field is computed by Schwinger. This is an exact one-loop 
nonperturbative result \cite{schwinger}. This result can also be understood 
in terms of
a semiclassical tunneling across the mass gap \cite{casher}. However, 
a space-time dependent field $A$ can directly excite the negative energy 
particles to levels above the mass gap,
by a perturbative mechanism $A \rightarrow q\bar q$, 
without recourse to any tunneling or barrier
penetration mechanism \cite{RajRavi}. The particle production from 
this mechanism is important in high energy heavy-ion collisions.
As shown by numerical studies
\cite{nayak}, the chromofield acquires a space-time dependence
as soon as it starts producing partons and so Schwinger's 
non-perturbative formula for particle production from a constant field
is not applicable. In this case the treatment of 
parton production from a space-time 
dependent chromofield is necessary.
It can be mentioned here that, so far the transport 
equation is solved with parton productions from a constant field taken 
into account \cite{all} because  gluon pair production from the space-time 
dependent chromofield is not computed. Aim of this paper is to
calculate the probability for the process $A \rightarrow gg$ which is
similar to the process $A \rightarrow q\bar q$.
Quark and gluon production from a space-time dependent
chromofield is needed to study the production and equilibration 
of a quark-gluon plasma in ultra relativistic heavy-ion collisions
at RHIC and LHC.

The production of $q \bar{q}$ pairs from a space-time dependent
non-abelian field is almost the same to that of the production of 
$e^{+}e^{-}$ pairs from the abelian field.
This is because, apart from the color factors, the interaction of 
the quantized Dirac field with the classical potential is the same
in both cases. All the methods used to obtain the probability
for the production of the $e^+e^-$ pair from a space-time dependent 
Maxwell field can be applied to obtain the probability for the 
production of the $q\bar{q}$ pair 
from a space-time dependent Yang-Mills field. On the other hand, 
the production of gluon pairs from a space-time dependent Yang-Mills field 
is not straight forward and there is no counter part to this in the 
abelian theory. In contrast to abelian theory the quantized Yang-Mills 
field interacts with the classical non-abelian potential. Because of 
this interaction there is gluon production from the QCD vacuum in 
the presence of an external chromofield. This phenomenon is absent in the 
Maxwell theory. 

We again mention here that the fermion pair production from the space-time
dependent field is studied in the literature \cite{schwinger,izju,RajRavi}
but gluon production from space-time dependent chromofield is 
not studied so far. This is because a consistent theory involving 
the interaction of the gluons with the classical chromofield is not available
in the conventional theory of QCD. In the case of fermions 
one knows exactly the theory for the interaction of fermions with 
the classical field. This is given by the Dirac equation (see section II). 
However, a consistent theory of gluons in the presence of external 
classical chromofield is not available in the conventional QCD.
This problem is addressed properly within the background
field method of QCD which was introduced by DeWitt and 't Hooft.
We will briefly mention the main differences between the background 
field method of QCD and the conventional method of QCD in section III. 
Unlike in conventional QCD, the Feynman diagrams obtained in 
the background field method contain classical chromofields and gluons. 
We will use the Feynman rules derived in the 
background field method of QCD to obtain the probability for
the processes $A \rightarrow q\bar q, gg$ via vacuum polarization.
The study of the production and evolution of the quark-gluon
plasma at RHIC and LHC by solving the relativistic non-abelian transport 
equation \cite{nayak} with parton production from a
space-time dependent chromofield will be undertaken in the future.

This paper is organized as follows. We compute the probability for the
process $A \rightarrow q\bar q$
in section II. The probability for the process $A \rightarrow gg$ 
is computed in section III. A brief description
about future research and conclusions can be found in section IV. 

\section{Probability for the process $ A_{cl} \rightarrow Q\bar{Q}$}

The production of $q\bar{q}$ pairs from a non-abelian field 
via vacuum polarization is simillar to that
of the production of $e^+e^-$ pairs from the abelian field. 
This is because the interaction lagrangian of the quantized Dirac field with
the classical gauge potential is similar in both the cases.
The amplitude for the lowest order process $A \rightarrow e^{+}e^{-}$ 
which contributes to the production of $e^+e^-$ pair 
from a space-time dependent 
classical abelian field $A^{\mu}$ is given by \cite{RajRavi}
\be
M=<k_1,k_2|S^{(1)}|0>=-ie\bar{u}(k_1)\gamma_{\mu}A^{\mu}(K)v(k_2).
\label{eamp}
\ee
In the above expression $k_1$, $k_2$ are the four momenta of the produced 
electron and positron respectively and $A(K)$ is the Fourier transform of 
the space-time dependent field $A^{\mu}(x)$ with $K=k_1+k_2$. 

The probability for pair production is:
\be
W^{(1)}= \int \frac{d^3k_1}{{(2\pi)}^3 2 k_1^0} \int 
\frac{d^3k_2}{{(2\pi)}^3 2 k_2^0} ~ \int d^4K ~
{(2 \pi)}^4 ~\delta^{(4)}(K-k_1-k_2) ~ T
\label{prob}
\ee
where 
\be
T=\Sigma_{spin}|M|^2.
\label{msq}
\ee
Performing the spin sum one obtains \cite{izju}
\bea
W_{e^+e^-}^{(1)}=\frac{\alpha}{3} \int_{K^2>4m_e^2} d^4K ~ 
{(1-\frac{4m_e^2}{K^2})}^{1/2}(1+\frac{2m_e^2}{K^2}) ~
[K\cdot A(K)K\cdot A(-K) \nonumber \\
-K^2A(K)\cdot A(-K)].
\label{izyks}
\eea
In the above expression $\alpha=\frac{e^2}{4\pi}$
is the coupling constant. Using
\be
F^{\mu \nu}(K)=-i(K^{\mu}A^{\nu}-K^{\nu}A^{\mu})
\label{fmunu}
\ee
one obtains
\be
W_{e^+e^-}^{(1)}=\frac{\alpha}{3} \int_{K^2>4m_e^2} d^4K ~
{(1-\frac{4m_e^2}{K^2})}^{1/2}(1+\frac{2m_e^2}{K^2}) ~
[{|E(K)|}^2-{|B(K)|}^2].
\label{sch}
\ee
This is the probability for the production of the $e^{+}e^{-}$ pair from 
a space-time dependent abelian field. This result was obtained for the 
first time by Schwinger (see Eq. (6.33) in \cite{schwinger}).

Now we proceed to compute the probability for the production of 
a $q\bar{q}$ pair from a space-time dependent non-abelian field.
The amplitude for the lowest order process $A \rightarrow q\bar{q}$
which contributes to $q\bar{q}$ pair production from a space-time dependent
classical non-abelian field $A^{a\mu}$ is given by
\be
M=ig\bar{u}^i(k_1)\gamma_{\mu}T^a_{ij}A^{a\mu}(K)v^j(k_2).
\label{qamp}
\ee
In the above expression 
$k_1$ and $k_2$ are the four momenta of the quark and anti-quark respectively
and $A^{a\mu}(K)$ is the Fourier
transform of the space-time dependent non-abelian field $A^{a\mu}(x)$ with
$K=k_1+k_2$. $T^a_{ij}$ are the generators in the fundamental
representation, $i,j$ are the color indices for the quarks and
$a$ is the color index of the non-abelian field. 

Following the above procedure and choosing SU(3) gauge group one obtains
\bea
W_{q\bar{q}}^{(1)}= tr[T^aT^b]\frac{\alpha_s}{3} \int_{K^2>4m_q^2} d^4K ~
{(1-\frac{4m_q^2}{K^2})}^{1/2}(1+\frac{2m_q^2}{K^2}) ~
[K\cdot A^a(K)K\cdot A^b(-K) \nonumber \\
-K^2A^a(K)\cdot A^b(-K)].
\label{izyks1}
\eea
In the above expression $\alpha_s=\frac{g^2}{4\pi}$. 
Using $tr[T^aT^b]=\frac{1}{2}\delta^{ab}$ one obtains
\bea
W_{q\bar{q}}^{(1)}=\frac{\alpha_s}{6} \int_{K^2>4m_q^2} d^4K ~
{(1-\frac{4m_q^2}{K^2})}^{1/2}(1+\frac{2m_q^2}{K^2}) ~
[K\cdot A^a(K)K\cdot A^a(-K) \nonumber \\
-K^2A^a(K)\cdot A^a(-K)].
\label{probq}
\eea
This is the probability for the production of a $q\bar{q}$ pair from a 
space-time dependent chromofield to the lowest order in the coupling constant 
$\alpha_s$. It can be noted that Eq. (\ref{probq}) 
is valid only for a single flavor of quarks. This equation is simillar to 
Eq. (\ref{izyks}) in the abelian theory. 

\section{Probability for the process $A \rightarrow gg$}

Now we proceed to compute the probability for the process 
$A_{cl} \rightarrow gg$ via vacuum polarization.
The amplitude for this process (see Fig. 1(a)) is given by
\be
M=\epsilon^{b\nu}(k_1)V^{abc}_{\mu \nu \lambda}(K,k_1,k_2)A^{a\mu}(K)
\epsilon^{c\lambda}(k_2).
\label{gamp}
\ee
In the above expression $k_1$, $k_2$ are the four momenta of the 
produced gluons, $A^{a\mu}(K)$ is the Fourier transform of the non-abelian
field $A^{a\mu}(x)$ and $V^{abc}_{\mu \nu \lambda}(K,k_1,k_2)$ 
is the vertex involving a single classical field and two gluons with
$K=k_1+k_2$. In the conventional method of QCD such vertices which involve 
classical field and gluons are not calculated. Hence, the gluon production
from a space-time dependent chromofield is not available. 
However, such a calculation is possible in the background field
method of QCD which was introduced by DeWitt and 't Hooft \cite{dewitt,thooft}. 
This is because the Feynman diagrams involving a classical chromofield 
and gluons are obtained in the background field method of QCD.
We use the Feynman rules obtained by the background field method of QCD to
compute the probability for the process $A_{cl} \rightarrow gg$ via vacuum
polarization. First of all we will briefly describe the main differences 
between conventional QCD and the background field method of QCD before 
studying the above process.

In the conventional theory of QCD the generating function is 
\be
Z[J]=\int [dA] ~det M_G ~exp(i [S[A]-\frac{1}{2\alpha}
G\cdot G + J \cdot A]),
\label{generc}
\ee
where we have not included the quark part for simplicity. In the above 
expression the gauge field action is
\be
S[A]=-\frac{1}{4}~\int d^4x ~{(F^{a\mu \nu})}^2
\ee
with
\be
F^{a\mu \nu} = \partial^{\mu}{A^{a\nu}}-\partial^{\nu}
{A^{a\mu}}+g~f^{abc}~{A^{b\mu}} ~{A^{c\nu}}
\ee
for the group with structure constants $f^{abc}$. The other two terms are
\be
J\cdot A = \int d^4x ~J_{\mu}^aA^{a\mu}
\ee
and
\be
G\cdot G = \int d^4x G^aG^a
\ee
with $G^a$ being the gauge fixing term. A typical choice for the
gauge fixing term in QCD is
\be
G^a=\partial_{\mu}A^{a\mu}.
\label{gf}
\ee
The matrix element of ${(M_G(x,y))}$ is given by
\be
{(M_G(x,y))}^{ab}=\frac{\delta(G^a(x))}{\delta \theta^b (y)}
\ee
where $\frac{\delta(G^a(x))}{\delta \theta^b (y)}$ is the derivative 
of the gauge fixing term under the infinitesimal gauge transformation
\be
\delta A^{a\mu}= -f^{abc}~\theta^b {A^{c\mu}}
~+~\frac{1}{g}~\partial^{\mu}\theta^a.
\ee
The $detM_G$ in the generating functional is 
written as functional integral over the {\it Faddeev-Popov ghost}~field 
$\chi^a$.
We mention here that any physical quantity calculated in this method is gauge 
invariant and independent of the particular gauge chosen. However, 
there is still a problem of gauge invariance with some of the quantities
like off-shell Green's functions or divergent counter terms. This problem
arises because in order to quantize the theory one must fix a gauge. This
means that the total lagrangian we actually use in conventional 
QCD, consisting of the classical lagrangian (which is explicitly
gauge invariant) plus a gauge-fixing and ghost terms, 
is not gauge invariant. The background field method is a technique which allows
one to fix a gauge in quantizing the theory without losing explicit
gauge invariance which is present at the classical level of the gauge
field theory. In the background field approach, one arranges things
so that explicit gauge invariance is still present once gauge fixing and
ghost terms are added. In this way Green's functions obey the naive Ward
identities of gauge invariance and the unphysical quantities like 
divergent counterterms becomes gauge invariant. However, for our purpose
of gluon productions from the space-time dependent chromofield 
to the leading order we do not need to address all these issues here.
The details about the explicit gauge invariance of QCD in the presence of
background chromofield can be found in \cite{dewitt,thooft,abbott,all1}. 
In our study we require Feynman rules involving the classical chromofield and 
gluons. For this reason we briefly outline the procedure of
background field method in QCD and present the generating functional 
which generates the Feynman rules involving gluons, ghosts and the
classical chromofields.

In the background field method of QCD the gauge fixing term is given 
by \cite{thooft}
\be
G^a=\partial^{\mu}A_q^{a\mu}+g~f^{abc}~A_{cl}^{b\mu}A_q^{c\mu},
\label{gfb}
\ee
which depends on the classical background field $A_{cl}^{a \mu}$. 
The variable of the integration in the functional integral is
the quantum gauge field $A_q^{a\mu}$ and, following 't Hooft \cite{thooft} 
the background field is not coupled to the external source $J$.
The generating functional depends on $J$ and $A_{cl}$ and is given by
\be
Z[J,A_{cl}]=\int [dA_q] ~det M_G ~exp(i [S[A_q+A_{cl}]-\frac{1}{2\alpha}
G\cdot G + J \cdot A_q]).
\label{gener}
\ee
The matrix element of $M_G$ is given by 
\be
{(M_G(x,y))}^{ab}=\frac{\delta(G^a(x))}{\delta \theta^b (y)}
\ee
where $\frac{\delta(G^a(x))}{\delta \theta^b (y)}$ is the derivative 
of the gauge fixing term under the infinitesimal gauge transformation
\be
\delta A_q^{a\mu}= -f^{abc}~\theta^b {(A_q^{c\mu}+A_{cl}^{c\mu})}
~+~\frac{1}{g}~\partial^{\mu}\theta^a.
\ee
Writing $detM_G$ as functional integral over the ghost field one finds
\bea
Z[J,A_{cl},\xi,\xi^{*}]=\int [dA_q] ~[d\chi] ~[d\chi^{*}] ~exp(i 
[S[A_q+A_{cl}]+S_{ghost}-\frac{1}{2\alpha} G\cdot G + J \cdot A_q \nonumber \\
+ \chi^{*} \xi + \xi^{*} \chi]).
\label{generf}
\eea
where $\xi$ and $\xi^*$ are source functions for the ghosts and
\bea
S_{ghost}= -\int d^4x ~\chi^{\dagger}_a[\Box^2 \delta^{ab}~-~
g {\overleftarrow{\partial}}_{\mu}f^{abc}(A_{cl}^{c\mu}+A_q^{c\mu}) 
~+~gf^{abc}A^{c\mu}\vec{\partial}_{\mu}~ \nonumber \\
+~g^2f^{ace}f^{edb}
A^c_{cl\mu}(A_{cl}^{d\mu}+A_q^{d\mu})]\chi_b.
\eea

The Feynman rules \cite{abbott} for QCD in the presence of a classical 
background chromofield are constructed from the generating functional 
which is given by Eq. (\ref{generf}) with the gauge fixing term given by 
Eq. (\ref{gfb}). We use the Feynman rules obtained in the background 
field method to compute the probability for the process 
$A_{cl} \rightarrow gg$.
Within this method the vertex $V^{abc}_{\mu \nu \lambda}(K,k_1,k_2)$, 
involving a single classical chromofield and two gluons which appear in 
the amplitude (See Fig. 1(a) and Eq. (\ref{gamp})) in Feynman gauge 
is given by
\be
V^{abc}_{\mu \nu \lambda}(K,k_1,k_2)= g f^{abc} ~[2g_{\mu\nu}K_{\lambda}-
2g_{\mu\lambda}K_{\nu}-g_{\nu\lambda}{\it k}_{\mu}],
\label{vertex}
\ee
where $K=k_1+k_2$ and $k=k_1-k_2$. The above vertex is different from the
three gluon vertex usually used in conventional QCD.

Using Eq. (\ref{gamp}) in (\ref{msq}) we find
\be
T_{gl}= \frac{1}{4}\Sigma_{spin} A^{a\mu}(K)A^{a^{\prime}{\mu}^{\prime}}(-K)
V^{abc}_{\mu \nu \lambda}(K,k) 
V^{*a^{\prime}b^{\prime}c^{\prime}}_{\mu^{\prime}
\nu^{\prime} \lambda^{\prime}}(K,k)
\epsilon^{b\nu}(k_1) \epsilon^{*b^{\prime}\nu^{\prime}}(k_1)
\epsilon^{c\lambda}(k_2) \epsilon^{*c^{\prime}\lambda^{\prime}}(k_2),
\label{trans}
\ee
where we have used $A^*(K)=A(-K)$. The factor $\frac{1}{4}$ is the 
weight factor given in the Feynman amplitude in order to obtain a
correct gauge field and gauge-fixing action.
To obtain correct and physical results we have to use the appropriate
projection operators for the transverse polarization states of the gluons.
For this purpose we proceed as follows. First of all we use the 
polarization sum
\be
\Sigma_{spin}\epsilon^{\nu}_1 \epsilon^{*\nu^{\prime}}_1 =
\Sigma_{spin}\epsilon^{\nu}_2 \epsilon^{*\nu^{\prime}}_2 = -g^{\nu \nu^{\prime}}
\label{spin1}
\ee
and then substract the corresponding ghost contributions (see Fig. 1(b)).

Using Eq. (\ref{vertex}), (\ref{spin1}) and performing the calculation 
we find
\bea
T_{gl}=\frac{1}{4}&&A^{a\mu}(K)A^{a^{\prime}{\mu}^{\prime}}(-K)
V^{abc}_{\mu \nu \lambda}(K,k)
V^{*a^{\prime}bc}_{\mu^{\prime} \nu^{\prime} \lambda^{\prime}}(K,k)
g^{\nu \nu^{\prime}} 
g^{\lambda \lambda^{\prime}}=
Ng^2 ~[ 4k_1\cdot k_2A^a(K)\cdot A^a(-K) \nonumber \\
&&-3k_1\cdot A^a(K)k_2\cdot A^a(-K) 
-3k_2\cdot A^a(K) 
k_1\cdot A^a(-K) -k_1\cdot A^a(K)k_1\cdot A^a(-K)  \nonumber \\
&&-k_2\cdot A^a(K)k_2\cdot A^a(-K)],
\label{t1}
\eea
where we have used $f^{abc}f^{a^{\prime}bc} =N\delta^{aa^{\prime}}$ 
in SU(N) gauge group. 

Now we proceed to compute the probability for the process 
$A_{cl} \rightarrow gg$. To obtain the probability
for the above process 
we require that $K^2={(k_1+k_2)}^2>0$ with $K^0>0$. We recall that for real 
gluons $k_1^2=k_2^2=0$. With the above requirements we proceed to perform
the integration in Eq. (\ref{prob}) using $T_{gl}$ from 
Eq. (\ref{t1}). As gluons are similar particles we multiply a factor
$\frac{1}{2}$ in the phase space and find
\bea
W_{gl}^{(1)}=\frac{1}{4}\frac{Ng^2}{32 \pi^2} \int_{K^2>0} d^4K ~\theta(K^0) 
\int \frac{d^3k_1}{k_1^0} \frac{d^3k_2}{k_2^0} \delta^{(4)}(K-k_1-k_2) 
A^a_{\mu}(K)A^a_{\nu}(-K) \nonumber \\
{[8g^{\mu \nu}K^2-8K^{\mu}K^{\nu}+4k^{\mu}k^{\nu}]} 
\label{wgg}
\eea
where $k=k_1-k_2$.

Now, we evaluate the corresponding ghost diagrams which have to be substracted
from the above result. The amplitude for the ghost part (see Fig. $1_b$) 
is given by
\be
M_{gh}^{bc}= g f^{abc} k^{\mu} A^a_{\mu}.
\label{gham}
\ee
so that
\be
T_{gh}= \frac{1}{4}M_{gh}^{bc}~ {M_{gh}^{bc}}^*
\label{ght}
\ee

Using Eq. (\ref{ght}) in (\ref{prob}) we find
\bea
W_{gh}^{(1)}=\frac{1}{4}\frac{Ng^2}{16 \pi^2} \int_{K^2>0} d^4K ~\theta(K^0) 
\int \frac{d^3k_1}{k_1^0} \frac{d^3k_2}{k_2^0} \delta^{(4)}(K-k_1-k_2) 
A^a_{\mu}(K)A^a_{\nu}(-K) {[k^{\mu}k^{\nu}]} 
\label{wgh}
\eea

Substracting Eq. (\ref{wgh}) from (\ref{wgg}) we find
\bea
W_{A \rightarrow gg}^{(1)}=\frac{1}{4}\frac{Ng^2}{32 \pi^2} 
\int_{K^2>0} d^4K ~\theta(K^0) 
\int \frac{d^3k_1}{k_1^0} \frac{d^3k_2}{k_2^0} \delta^{(4)}(K-k_1-k_2) 
A^a_{\mu}(K)A^a_{\nu}(-K) \nonumber \\
{[8g^{\mu \nu}-8K^{\mu}K^{\nu}+2k^{\mu}k^{\nu}]} 
\label{wgl}
\eea

To perform the integration in the above equation we proceed as follows.
First of all using
\be
\int \frac{d^3k_2}{2 k_2^0} =
\int d^4k_2 ~\delta(k_2^2) ~\theta(k_2^0)
\label{fourk}
\ee
we check that
\be
\int \frac{d^3k_1}{k_1^0} \frac{d^3k_2}{k_2^0} \delta^{(4)}(K-k_1-k_2)
g^{\mu\nu} = 2 \pi g^{\mu\nu}.
\label{id1}
\ee
Using Eq. (\ref{id1}) we find
\bea
\int \frac{d^3k_1}{k_1^0} \frac{d^3k_2}{k_2^0} \delta^{(4)}(K-k_1-k_2) 
k^{\mu}_1 k^{\nu}_2 = 
\int \frac{d^3k_1}{k_1^0} \frac{d^3k_2}{k_2^0} \delta^{(4)}(K-k_1-k_2) 
k^{\mu}_2 k^{\nu}_1 \nonumber \\
= \frac{\pi}{6}[K^2g^{\mu\nu}+2K^{\mu}K^{\nu}], 
\label{id2}
\eea
and
\bea
\int \frac{d^3k_1}{k_1^0} \frac{d^3k_2}{k_2^0} \delta^{(4)}(K-k_1-k_2) 
k^{\mu}_1 k^{\nu}_1 = \int \frac{d^3k_1}{k_1^0} \frac{d^3k_2}{k_2^0} 
\delta^{(4)}(K-k_1-k_2) k^{\mu}_2 k^{\nu}_2 \nonumber \\
= \frac{\pi}{6}[-K^2g^{\mu\nu}
+4K^{\mu}K^{\nu}]. 
\label{id3}
\eea
With the help of the above relations the integration in Eq. (\ref{wgl})
can be easily performed. Using (\ref{id1}), 
(\ref{id2}) and (\ref{id3}) we obtain from Eq. (\ref{wgl}) 
\be
W_{Agg}^{(1)}=\frac{11 N \alpha_s}{24} 
\int_{K^2>0} d^4K ~\theta(K^0) 
~[K^2A^a(K)\cdot A^a(-K)-K\cdot A^a(K)K\cdot A^a(-K)]
\label{probgn}
\ee
In the above expression the repeated indices are summed from 
$a=1,...(N^2-1)$. For SU(3) gauge group we obtain
\be
W_{Agg}^{(1)}=\frac{11 \alpha_s}{8} \int_{K^2>0} d^4K ~\theta(K^0)~[
K^2A^a(K)\cdot A^a(-K)-K\cdot A^a(K)K\cdot A^a(-K)]
\label{probga}
\ee
This is the probability computed from the 
process $A \rightarrow gg$ via vacuum polarization. 
It can be seen that the above
expression is transverse with respect to the momentum of the field
$K$, i.e, when $A^{\mu}$ is replaced by $K^{\mu}$ we get
$W_{Agg}^{(1)} =0$. This result is a part of the expression
for the total probability (including higher order terms in $gA$)
for the production of gluon pairs.

\section{Conclusions}

Using the background field method of QCD, we have computed the probability
for the processes $A \rightarrow q\bar q, gg$ via vacuum polarization,
in the presence of a space-time dependent chromofield $A$.
These processes are similar to $A \rightarrow e^+e^-$ in QED.
For any arbitary classical non-abelian chromofield  
one has to check the gauge invariance of the result with respect to
the general non-abelian local gauge transformation which is known
as type I gauge transformation \cite{itz}.
Under this gauge transformation the classical chromofield transform like:
\be
A^{\prime \mu}_{cl} \rightarrow UA^{\mu}_{cl} U^{-1} - \frac{i}{g}
(\partial_{\mu} U) U^{-1}.
\label{gtc}
\ee
The gluonic and ghost fields transform like:
\be
A^{\prime \mu}_{q} \rightarrow UA^{\mu}_{q} U^{-1}, 
\label{gtg}
\ee
and
\be
\chi^{\prime} \rightarrow U\chi U^{-1} 
\label{gtgh}
\ee

respectively.

In these situations one has to add the results of the higher order 
processes (for the gluon pair production)
to the leading order results obtained in this paper. 
The gluon pair production amplitude which is obtained from the
first order $S$ matrix contains higher order terms in $gA$. This is
due to the non-abelian nature of the gluonic and classical
chromofield. The result of the gluon pair production probability
which contains both leading order and higher order terms in $gA$
(but first order in $S$ matrix) will be reported elsewhere \cite{higher}. 
In this paper we have presented only the
result from the leading order processes $A \rightarrow gg$.
This result is a part of the expression for the total probability
of the gluon production from a space-time dependent chromofield \cite{higher}.
Quark and gluon production from a
space-time dependent chromofield will play a crucial role in the
production and equilibration of the quark-gluon plasma in ultra relativistic
heavy-ion collsions. Our main goal is to
study the production and equilibration of the
quark-gluon plasma by solving relativistic non-abelian transport
equations for partons (see Eq. (\ref{trans1})) with parton production 
from a space-time dependent chromofield taken into account.
Such work requires extensive numerical computation \cite{nayakb}
and needs a separate publication.

\acknowledgements

We thank Dennis D. Dietrich for useful discussions, reading of the
manuscript and drawing the Feynman diagrams. We also thank Dr. Qun Wang
and Dr. Chung-Wen Kao for useful discussions. G.C.N. acknowledges the financial 
support from Alexander von Humboldt Foundation.




\begin{figure}[thb]
\begin{center}
\pfig{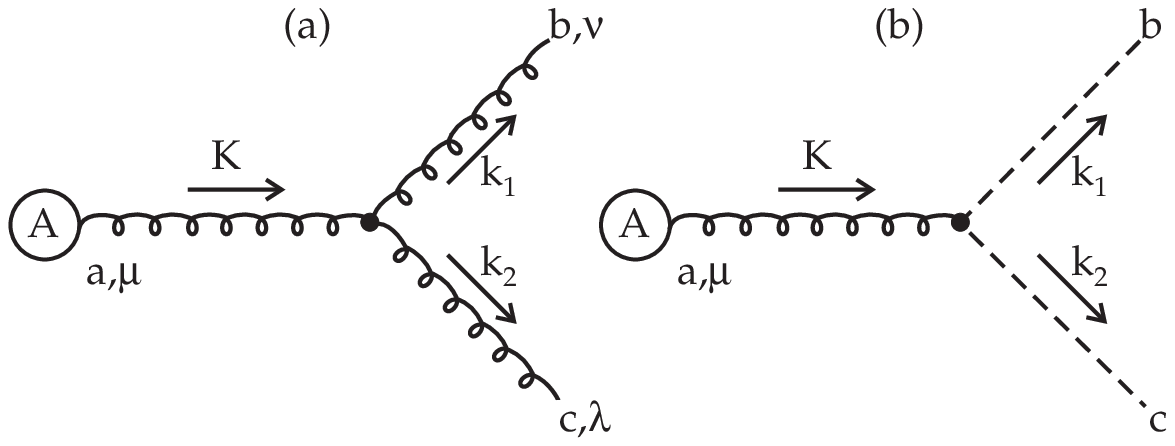}{12cm}
{Lowest order vacuum polarization diagrams. Fig. 1(a) is the gluon 
diagram and Fig. 1(b) is the corresponding ghost diagram.}
\end{center}
\end{figure}

\end{document}